# STATUS OF THE WARM FRONT END OF PXIE

A. Shemyakin[#], M. Alvarez, R. Andrews, C. Baffes, A. Chen, B. Hanna, L. Prost, G. Saewert, V. Scarpine, J. Steimel, D. Sun, Fermilab*, Batavia, IL 60510, USA
D. Li, LBNL, Berkeley, CA 94720, USA
R. D'Arcy, UCL, London, UK, WC1E 6BT, UK and Fermilab

## Abstract

A CW-compatible, pulsed H- superconducting linac is envisaged as a possible path for upgrading Fermilab's injection complex. To validate the concept of the front-end of such a machine, a test accelerator (a.k.a. PXIE) is under construction. The warm part of this accelerator comprises a 10 mA DC, 30 keV H- ion source, a 2m-long LEBT, a 2.1 MeV CW RFQ, and a 10-m long MEBT that is capable of creating a large variety of bunch structures. The paper will report commissioning results of a partially assembled LEBT, status of RFQ manufacturing, and describe development of the MEBT, in particular, of elements of its chopping system.

## INTRODUCTION

A proposal to upgrade Fermilab's injection complex suggests building a CW SRF H- linac, which at its initial stage, known as the Proton Improvement Plan –II (PIP-II) [1], will be used in a pulsed mode. Front end components crucial for CW operation will be tested at an accelerator called, for historical reasons, PXIE [2]. PXIE consists of an H- ion source; a Low Energy Beam Transport (LEBT); a CW 2.1-MeV RFQ; a MEBT; two SRF cryomodules (HWR and a SSR1); a High Energy Beam Transport (HEBT); and a beam dump (Fig.1). This paper describes the status of various components.

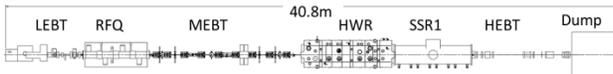

Figure 1: PXIE beamline layout.

## LOW ENERGY BEAM TRANSPORT

The LEBT [3] is being commissioned in a straight configuration as depicted in Fig.2 (eventually, it will include a dipole magnet bend). It consists of an H- volume-cusp ion source [4], 3 solenoids, a chopper assembly and diagnostics.

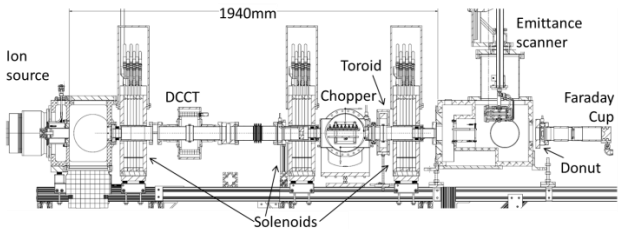

Figure 2. Mechanical schematic of the LEBT.

Each solenoid has a pair of built-in dipole correctors. Vacuum chambers of the first two solenoids contain electrically isolated diaphragms to control the beam neutralization and estimate beam position in combination with the correctors. The beam current is monitored by a DCCT, a current transformer (toroid), and a Faraday cup at the end of the line. Varying the potential of the ion source extraction electrode creates pulsed beam with a pulse length of 0.01 ms – 16 ms at 10 Hz.

All LEBT components, except the chopper, have been recently installed and are operational. Chopper installation is planned for the fall of 2014.

Transverse emittance is measured by an Allison emittance scanner, developed in collaboration with SNS. The scanner operates in both pulse and DC modes. It was commissioned with a one-solenoid LEBT configuration, and some results are shown in Fig. 3.

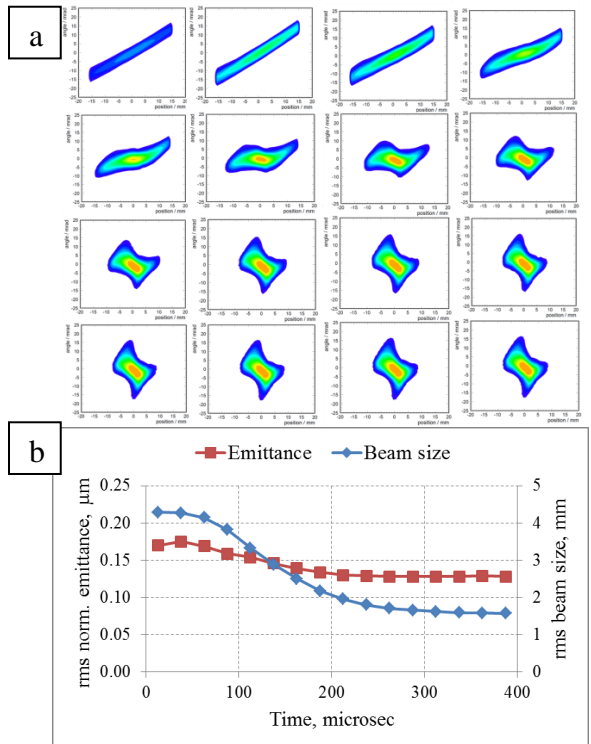

Figure 3: Evolution during a 0.4 ms, 5mA beam pulse. a- phase portraits in 25 μs temporal bins. Vertical axis is ±25mrad, horizontal ±20 mm. The same color represents the same phase density on all plots. b- rms, norm. beam emittance and rms beam size for the same set.

The measured beam emittance is within the ion source specifications and, at low currents, is in agreement with the ion source acceptance measurements, ~0.1 µm (rms norm.). However, emittance rises rapidly as a function of beam current above 4 mA and reaches 0.25 µm for 9 mA. Higher intensity optimization is on-going.

Emittance is also measured using an isolated diaphragm in front of the Faraday cup ("donut"). Analysis of the current lost at the donut, as beam is moved across it, allows reconstructing the beam size. Beam size vs. upstream solenoid current is fitted to a simple, free space model. Space charge effects are assumed negligible for DC or long-pulse modes. In the one-solenoid configuration, this method agrees with Alison scanner data.

Preliminary results in the 3-solenoid scheme show no significant emittance growth in comparison with the one – solenoid assembly. A detailed description of all emittance measurements is presented in [5].

The LEBT optics is being measured by monitoring beam loss in the donuts vs. dipole kicks in various locations. We will continue to measure the beam envelope and perform experiments with partially neutralized transport to be ready for commissioning the RFQ.

## RFQ

The RFQ design is for 5 mA $H^-$ CW operation with 2.1 MeV output energy [6]. The structure consists of four 1.1m, 4-vane RF modules with pi-mode rods and operates at the resonant frequency at 162.5 MHz. The RFQ has been designed by LBNL in collaboration with Fermilab. The RFQ is being fabricated at LBNL: machining of all 16 vanes is nearly complete; Module #2 has been assembled, brazed and is currently being vacuum-leak checked. RF bead-pull measurements of Module #2 were performed before brazing using temporary cut-backs and 20 temporary tuners set to nominal 20-mm intrusion to the cavity body. The measurements agree with simulations and show the required field flatness (Fig. 4 from [7]). The frequency difference between simulation and measurement was only 160 kHz. The RFQ delivery to Fermilab is expected in April 2014.

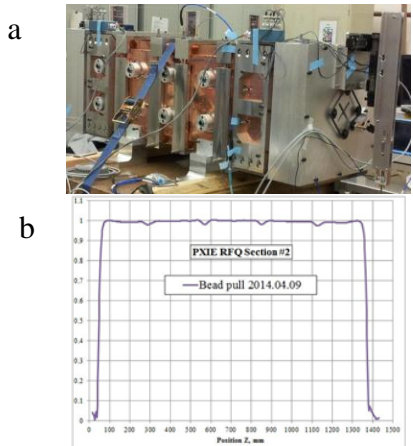

Figure 4: Bead pull measurements of the Module 2. a- assembly; b- normalized field flatness (average of 4 quadrants).

Two RF power couplers [8] are being constructed and will be tested in the fall of 2014. It is worth pointing out that an RFQ of a very similar design by LBNL was successfully commissioned recently at Institute of Modern Physics (IMP) (Lanzhou, China) [9].

The RFQ will be powered by two 75 kW solid-state amplifiers from SigmaPhi. The amplifiers have been installed outside the PXIE enclosure and are being commissioned.

## MEBT

The MEBT design is described in Ref.[10]. Transverse and longitudinal focusing in the MEBT will be provided by 25 quadrupoles combined into doublets and triplets, and by three 162.5 MHz bunching cavities, correspondingly (Fig.5). The quadrupoles have been designed and will be fabricated at BARC in India. A prototype cavity is being manufactured in the US industry, and high-power testing is expected in 2014. A concept of a radiation –cooled scraper was tested with an electron beam [11], and a prototype scraper assembly is in production.

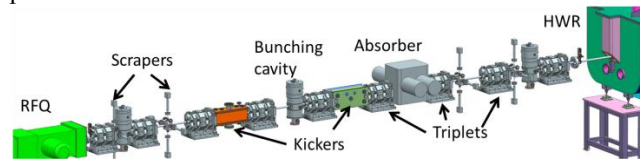

Figure 5. 3D model of MEBT. Diagnostics are not shown.

The most challenging part of the MEBT design is the chopping system that forms the required bunch pattern by deflecting undesired bunches of the initially CW beam into an absorber. This deflection is made by two broadband, travelling-wave kickers separated by 180º of the betatron phase advance and operating synchronously. Details of the scheme are in Ref. [12]. Two versions of the kicker, distinguished by the characteristic impedance of the kicker structure (50 and 200 Ohm), are being developed.

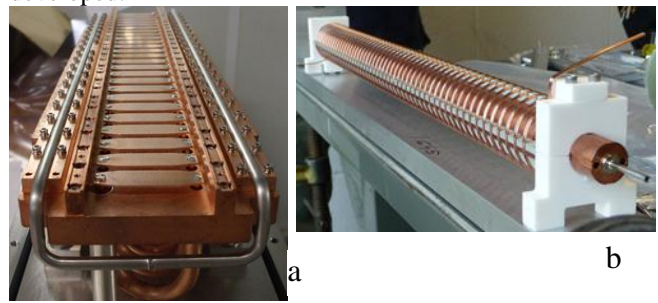

Figure 5: MEBT kicker. a – plate of 50 Ohm kicker before installation into the vacuum box. b – model of the 200 Ohm kicker.

In the 50 Ohm version, the beam is deflected by voltage applied to planar electrodes connected in vacuum by coaxial cables providing the necessary delays (Fig. 5a). One plate of the kicker prototype was assembled in its final configuration. Low power RF measurements showed the expected dispersion and power loss characteristics.

The average phase velocity deviates from the expected beam speed by less than 0.1%. This plate is under vacuum in the kicker box for high power testing. The plate was subjected in vacuum to a 550W, 250 MHz CW signal that created, according to simulation, the power loss exceeding the maximum loss expected at the nominal pulse characteristics. The preliminary results are encouraging. With full power, the plate electrode temperature rise measured with an infrared camera was ~15C, and the vacuum degradation was ~$5 \cdot 10^{-7}$ Torr and was decreasing with conditioning (with 100 l/s ion pump).

The 200 Ohm kicker consists of two helixes wound around grounded cylinders. While heating caused by RF is low in this version, effects of possible beam loss to electrodes can be significant. To address these concerns, the spacers between the helix winding and the water-cooled ground tube are made from high thermal conductivity AlN ceramics glued with vacuum-compatible epoxy (Fig. 5b).

Proposals for the kicker drivers in both versions are presented in [11]. There were no significant development efforts in FY14.

Both kicker prototypes are expected to be assembled in their complete configuration and tested to full heating in vacuum (accounting for both RF loss and possible beam irradiation of 40W per plate) in FY15. If tests are successful, the kickers will be installed into MEBT in FY16 to measure the beam response, with the plan to demonstrate bunch-by-bunch selection in FY17.

Another part of the chopping system is the absorber, which concept is described in Ref. [13]. Thermal features of the concept were successfully tested with two short (~1/4 size) prototypes irradiated by an electron beam with the representative absorbed power density [14]. The second prototype is presently under thermocycling tests and is expected to be tested with H- beam from RFQ in FY15.

## PLANS

The LEBT will be fully commissioned by the time of the RFQ arrival in spring of 2015. RFQ commissioning will be followed by beam characterization using mostly elements of the future MEBT. A 3D model of one of possible configurations is shown in Fig. 6. In part, it will include a toroid identical to LEBT's with intention to measure the beam loss in RFQ at the 1% level.

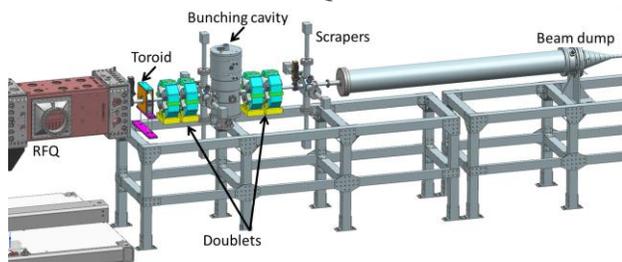

Figure 6: 3D model of the assembly for RFQ beam characterization. Image of doublets is courtesy of BARC.

A full-length MEBT is planned to be assembled in FY2016, initially with prototype elements for the chopping system, and the warm front end is expected to be finalized in FY2017 to be ready for installations of cryomodules.

## ACKNOWLEDGMENT

The authors would like to thank B. Brooker, K. Carlson, J. Czajkowski, M. Kucera, D. Snee, T. Zuchnik and their teams who built the LEBT beam line. We also would like to acknowledge the help we received from Dr. Qing Ji, from LBNL, who was instrumental in the successful start-up of the ion source at Fermilab.